\documentclass[aps,prb,nofootinbib,unsortedaddress,a4paper
]{revtex4}
\usepackage{graphicx}
\usepackage{bm}

\begin{document}

\title{Structure and stability of chiral beta--tapes: \\ 
a computational coarse--grained approach}

\author{Giovanni Bellesia%
\def\thefootnote{\alph{footnote})}%
\footnote{Electronic address: Giovanni.Bellesia@ucd.ie}
\affiliation{Theory and Computation Group, Department of Chemistry, 
University College Dublin, Belfield, Dublin 4, Ireland}
}

\author{Maxim V. Fedorov%
\def\thefootnote{\alph{footnote})}%
\footnote{Electronic address: Maxim.Fedorov@ucd.ie}
}
\affiliation{Theory and Computation Group, Centre for Synthesis and Chemical Biology,
 Conway Institute for Biomolecular and Biomedical Research, Department of Chemistry,
 University College Dublin, Belfield, Dublin 4, Ireland}

\author{Yuri A. Kuznetsov%
\def\thefootnote{\alph{footnote})}%
\footnote{Electronic address: Yuri.Kuznetsov@ucd.ie}
}
\affiliation{Computing Centre,
University College Dublin, Belfield, Dublin 4, Ireland}

\author{Edward G. Timoshenko%
\def\thefootnote{\alph{footnote})}%
\footnote{Author to
whom correspondence should be addressed. Phone: +353-1-7162821,
Fax: +353-1-7162536, E-mail: Edward.Timoshenko@ucd.ie, 
Web page: http://darkstar.ucd.ie}
}
\affiliation{Theory and Computation Group, 
Centre for Synthesis and Chemical Biology,
Conway Institute of Biomolecular and Biomedical Research, 
Department of Chemistry, University College Dublin, Belfield, 
Dublin 4, Ireland}

\date{\today}

\begin{abstract}
We present two coarse--grained models of different levels of detail
for the description of beta--sheet tapes obtained from equilibrium 
self--assembly of short rationally designed oligopeptides in solution.
Here we only consider the case of the homopolymer oligopeptides with
the identical sidegroups attached, in which the tapes have a helicoid
surface with two equivalent sides.
The influence of the chirality parameter
on the geometrical characteristics, namely the
diameter, inter--strand distance and pitch, of the
tapes have been investigated. The two models are found to produce
equivalent results suggesting a considerable degree of universality
in conformations of the tapes.
\end{abstract}

\maketitle

\section{\label{sec:intro}Introduction}

Structured proteins in their folded state possess a rich variety
of three dimensional conformations ({\it tertiary structures})
which are usually classified in terms of the
mutual arrangements of the so--called motifs and
elements of the {\it secondary structure} \cite{tooze}.
The secondary structures are usually characterized 
as segments of the protein chain possessing a strong regularity in the 
values of the Ramachandran angles \cite{rama1,rama2}.
The most common of such structures are the $\alpha$-helix and the $\beta$-sheet.

There is a general view that the microscopic chirality of individual amino 
acids is responsible for the twisted shape of $\beta$-sheets in 
globular and fibrous proteins \cite{salemme}.
Apart from the flat conformation described by Pauling and Corey \cite{linus},
a $\beta$-strand can acquire a nonzero degree of helicity 
with a finite twist along the principal axis of the 
polypeptide chain. Furthermore, the helical structure of a single strand 
directly relates to the \emph{macroscopic} twisted 
shape of the whole $\beta$-sheet.

The $\beta$-sheet conformation has been recently exploited as a 
reference structure for the novel biomaterials 
produced by large--scale self--assembly of oligopeptides in solution
\cite{soms1,soms2,mit1,mit2}.
In the latter Refs.\ it has been shown that oligomeric peptides can be 
rationally designed so that they self--assemble unidirectionally in solution
forming helically twisted $\beta$-sheet tapes stabilized by a regular 
arrangement of hydrogen bonds. The main factors which stabilise the
tape structures are: the intermolecular hydrogen bonding between
the polypeptide backbones, cross--strand attractive forces between 
sidegroups (hydrogen bonding, electrostatic and hydrophobic) and 
lateral recognition (steric and $\pi$-$\pi$ interactions) 
between the adjacent $\beta$-strands.

The tape structure is regarded as only the 
first in the hierarchy of equilibrium structures observed 
with increasing oligopeptide concentration such as the double tapes, 
fibrils, fibers, and eventually nematic gels \cite{soms1,soms2,mit1,mit2}.

This novel route towards engineering of biomaterials is an alternative
to some approaches which caused controversy 
and provides a possibility for simple 
equilibrium control of the biomaterials architecture, their functional and 
mechanical  properties, as well as the kinetics of their formation
in response to pH, temperature and other triggers.
Among different applications of these biomaterials which could be envisaged
one can mention their current use as three--dimensional scaffolds for 
tissues growth \cite{mit1,mit2}.
Moreover, these oligopeptides--based assemblies serve as a simple
experimental model system which could be used for providing valuable insights 
into the self--assembly and aggregation mechanisms of natural proteins and, 
in particular, formation of plaques of $\beta$-amyloids \cite{amyl} and 
fibrous proteins structures.

In terms of computer simulations, a number of Molecular Dynamics studies
of oligopeptide systems were reported recently  \cite{mit1,soms3}. 
These works have explored the (meta)stability of relatively small 
aggregates over fairly short computing times accessible to simulations, thereby
providing valuable structural information which is difficult
to extract from the experiment. Unfortunately, it is difficult to
establish whether those structures were sufficiently well equilibrated
over such short run times and only several particular oligopeptide sequences
were considered.

At present, there is still no full understanding regarding the details of the
functional relationship between the chiral nature of the single $\beta$-strand
and the helical geometry of the $\beta$-tape.
More generally, only in recent years the connection 
between the molecular handedness and the morphology of supramolecular 
assemblies was examined \cite{selinger,harris,kornyshev}.
Researchers have found that chirality controls the shape 
of the macroscopic assemblies not only in natural and synthetic 
peptides, but also in other biological systems such as lipids, fatty acids, 
and nucleic acids (see Ref. \onlinecite{selinger} for a review on models 
for the formation of helical structures made up of chiral molecules).

Recently, Nandi and Bagchi \cite{nandi1,nandi2,bag} have proposed a model
for the assembly of chiral amphiphilic molecules.
The latter is based on a simplified representation of either the 
geometry or the potential energy, which is then minimized
in order to find the most efficient packing.
Their results, which are consistent with the experiment, show that chiral 
tetrahedral amphiphiles of the same handedness assemble at a finite 
angle with respect to their neighbors, driving the formation process of 
helical clusters.
In $\beta$-sheet tapes a similar behaviour is found, where the microscopic 
chirality arises from the intra--molecular interactions.
The intermolecular forces then stabilize the tape structure with
a finite twist angle observed between the neighboring strands.
This twist angle transfers the chirality from the single strands to the 
level of the mesoscopic assembly, which hence possesses chirality as well.
This type of the secondary structure could be rationalised as a
compromise between the \emph{out--of--plane} energy term originated 
from the chirality of the single peptides and the inter--molecular 
(mainly hydrogen bonding) energies of the backbones atoms, 
preferring a flat arrangement.

The main goal of this work is to achieve a fundamental understanding 
of the way in which the microscopic chirality of single peptide molecules 
manifests itself at a larger supramolecular scale of the self--assembled tapes.
In practice, we would like to construct a \emph{minimal} model capable of 
capturing the most essential features of this phenomenon.
For this, we shall adopt a simplified coarse--grained description for 
the rod--like oligopeptides with a nonzero degree of helicity. 
Our computational study will be based on classical Monte Carlo simulations in 
continuous space with the use of the standard Metropolis algorithm
\cite{metropolis} much used in polymer simulations.

To describe the inter--molecular hydrogen bonding 
occurring within two--dimensional $\beta$-tapes
we shall use a coarse--grained description via a combination of the 
soft--core repulsion terms and short--ranged attractive terms. 
Furthermore, the microscopic chirality is introduced via an \emph{ad hoc} 
quadratic term. The functional dependence of the macroscopic 
twist on the strength of the latter will then be analysed.
The explicit forms for different potential energy terms, including those 
describing the bonded interaction, and motivation for their choice
are detailed in the next section.


\section{\label{sec:meth}Methods}

A peptide $\beta$-sheet is a regular secondary structure 
characterised by values of the Ramachandran angles lying 
within the upper left quadrant of the Ramachandran plot \cite{tooze}.
Its basic units are short peptide segments (called $\beta$-strands) 
which are stabilized by an ordered network of hydrogen bonds 
between the atoms of the backbone. The spatial sequence of the 
$\beta$-strands can follow a \emph{parallel} or
\emph{antiparallel} pattern, depending on the reciprocal arrangement of 
the strands termini.

Whether $\beta$-sheets are formed by $\beta$-strands connected 
via loops regions within the same chain (intra--molecular sheets) 
or from many oligomeric peptides (inter--molecular sheets), as e.g.
in synthetic peptides assemblies and 
$\beta$-amyloids, they all conform to a variety of twisted and curved 
geometrical surfaces \cite{salemme}. The twisting appears in $\beta$-sheets 
due to the nonzero chirality of the single strands, the
backbone geometry of which can be well approximated
by a circular helix. The mathematical definition for the latter 
is given by the classical differential geometry of curves \cite{struik}.

While fully atomistic computational studies of proteins are quite common, 
there is also a tradition of using simplified off--lattice models 
in the literature \cite{levitt1,levitt2}. The latter facilitate simulations
by retaining the {\it most essential features} of the peptides and overcome
the difficulty with rather excessive equilibration times of the fully
atomistic systems, which often raise doubt as to the validity of the final
results for them. 
A number of important results on fast--folding proteins and peptides
have been obtained in the last decade via such coarse--grained approaches
\cite{thiru1,thiru2,takada,bag,marg1}.

In the search of a suitable simplified model for the $\beta$-strand capable of 
generating a stable two--dimensional tape, one can use the so--called 
$C_{\alpha}$ models \cite{marg2}, which retain one interaction site per residue.
Then the inter--molecular hydrogen bonds could be incorporated,
for example, via the angular--dependent effective potential \cite{klim}.

Another direction is to consider a different class of 
the so--called $C_{\alpha}$-$C_{\beta}$ intermediate level models 
\cite{marg2}, in which two or three interaction sites are retained
per residue in the backbone.
A two--dimensional aggregate can then be generated by
mimicking the hydrogen bonding via a sum of effective Lennard--Jones 
terms \cite{takada}, also incorporating the steric effects of the 
side chain.
Although this is still a crude representation for the hydrogen bond, we find 
the second route quite satisfactory for our purposes.

\subsection{Geometry of the simplified model}

In our model, the coarse--grained geometry of the single $\beta$-strand 
retains three interaction sites per residue.
The backbone of the single amino acid is represented by two beads named 
$C$ and $N$, standing for the moieties $C_{\alpha}HC'O$ and $NH$, 
respectively.
Each sidegroup (amino acid residue) is then modeled by a 
bead $S$ bonded to $C$.
One could also easily introduce many types of the sidegroups, but
we shall defer studying more complicated sequences to the future
publications until we are fully satisfied with the performance of
the simplest models of homogeneous sequences.

We shall introduce all of the energy parameters of our coarse--grained
models in units of $k_B T$. It should be noted, however, that these parameters
are {\it effective} as we have reduced the number of degrees of
freedom considerably by introducing united atoms and by describing 
the solvent implicitly.
Thus, these parameters are temperature--dependent, and so are the equilibrium
conformations, even though $k_B T$ cancels out formally from the Boltzmann
weight $\exp(-E/k_B T)$. In principle, one could determine how the 
coarse--grained parameters are related to the fully atomistic ones at a given
temperature via a procedure analogous to that of Ref. \onlinecite{btg}.
However, as any inverse problem, it is a considerably difficult task.

In practice, we have chosen the temperature equal to $300$ K and
the numerical values of most of the energy parameters so that they broadly 
correspond to the typical values in the fully atomistic force fields.
Concerning the purely phenomenological parameters, such as the chirality
parameter, their values were chosen so that a reasonable experimental range
of the twist in the structures is reproduced.

\subsection{Potential energy function of the model $A$}

The first choice for the potential energy model follows the guidelines of 
the model proposed by Thirumalai and Honeycutt 
\cite{thiru1,thiru2} in their mesoscopic simulations of $\beta$-barrels.
This \emph{minimal} forcefield model adopts functional forms
of interactions akin to those typically employed in fully atomistic 
molecular mechanics models.

\subsubsection{Bond length potential}
The length of each bond connecting two monomers is restrained towards 
the equilibrium value via a harmonic potential:
\begin{equation}
U_{bond}=\frac{K_{b}}{2}(r-r_{eq})^{2},
\end{equation}
in which $K_{b}=200.0\ k_{B}T\cdot\mbox{\AA}^{-2}$ and $r_{eq}=2.0\ \mbox{\AA}$.

\subsubsection{Bond angle potential}

Bond angles defined via triplets $C_{i}-N_{i+1}-C_{i+1}$, 
$N_i-C_i-N_{i+1}$,
$N_{i}-C_{i}-S_{i}$ and $S_{i}-C_{i}-N_{i+1}$ are controlled
via a harmonic potential of the form:
\begin{equation}
U_{angle}=\frac{K_{\theta}}{2}(\theta-\theta_{eq})^{2},
\end{equation}
where $K_{\theta}=40.0\text{ } k_{B}T$ and $\theta_{eq}=120^{\circ}$, 
$\theta_{eq}=0^{\circ}$ for angles centered at $C$ and $N$ 
respectively \cite{Baumketner}.

\subsubsection{Dihedral angle potential}

Dihedral angle $\alpha$ is defined by the following formula:
\begin{equation} \label{eq:dihe}
\alpha=\mbox{sign}(\alpha)\cdot \text{arccos}\Bigg(\frac{(\bm{r}_{12}\times\bm{r}_{32})\cdot(\bm{r}_{32}\times\bm{r}_{34})}
{\|\bm{r}_{12}\times\bm{r}_{32}\|\|\bm{r}_{32}\times\bm{r}_{34}\|}\Bigg),
\end{equation}
where $\bm{r}_{ij} = \bm{r}_{i}-\bm{r}_{j}$ and
\begin{equation}
\mbox{sign}(\alpha)=\mbox{sign}(\bm{r}_{12}\cdot\bm{r}_{32}\times\bm{r}_{34}).
\end{equation}
Torsional degrees of freedom are constrained by a sum of terms associated
with quadruplets of successive $C$ beads and having the form
\cite{Baumketner}:
\begin{equation}
U_{tors}=-A \cos(3\alpha)-B \cos(\alpha),
\end{equation}
where $A=B=4.0\text{ } k_{B}T$.

An additional dihedral term is introduced to force planarity 
between pairs of subsequent $S$ beads:
\begin{equation}
U_{plane}=D \cos(\alpha).
\end{equation}
It is applied to the quadruplets $S_{i}-C_{i}-C_{i+1}-S_{i+1}$ 
and $D=4.0\text{ } k_{B}T$.
The presence of this term, increases the stability of the structures, 
by enhancing the steric hindrance due to the side chains.
As mentioned above, this excluded volume effect is quite essential for
generating a two--dimensional tape as the inter--molecular 
hydrogen bonds have no directional dependencies in our model.

\subsubsection{Chirality}

Handedness is introduced in the model 
via a quadratic term involving only quadruplets of 
successive $C$ beads, that is:
\begin{equation}\label{eq:tau1}
U_{chiral}=\frac{K_{\tau}}{2}(\tau-\tau_{0})^{2},
\end{equation}
where $K_{\tau}= 10\,k_B\,T$
and $\tau$ is equal to the normalized numerator of the analogous 
quantity defined in Frenet--Serret picture of spatial curves
\cite{struik}:
\begin{equation}\label{eq:tau2}
\tau = \frac{\bm{r}_{12}\cdot\bm{r}_{22}\times\bm{r}_{34}}
{\|\bm{r}_{12}\|\|\bm{r}_{23}\times\bm{r}_{34}\|}.
\end{equation}
The dependence of the chirality parameter on the temperature is expected
to be relatively weak.
This is in qualitative correspondence with the experimental data \cite{soms1}, 
which has revealed high structural stability of tape assemblies,
in a wide range of temperatures, essentially while water remains liquid.
However, more experimental data is required in order to determine how exactly
$K_{\tau}/k_B T$ depends on the temperature.

\subsubsection{Non--bonded Interactions}

A short--ranged Lennard--Jones term is used here in order to represent, 
in a highly simplified way, the inter--molecular hydrogen bonding typical 
of $\beta$-sheet structures:
\begin{equation}
U_{LJ}=\epsilon_{1}\Big[\Big( \frac{\sigma^{LJ}}{r}\Big)^{12}
-\Big( \frac{\sigma^{LJ}}{r}\Big)^{6}\Big],
\end{equation}
where the energy constant is $\epsilon_{1}=5.0\text{ } k_{B}T$ for all 
the interactions involving the backbones beads.
The van Der Waals radius is taken as $\sigma^{LJ}=2.0\text{ }\cdot\mbox{\AA}$ 
for $C-C$ and $N-N$ inter--molecular interactions
and as $\sigma^{LJ}=3.0\text{ }\mbox{\AA}$ for $C-N$ pairs. A small attractive 
well is introduced between pairs of $S$ united atoms by
choosing $\epsilon_{2}=1.0\text{ } k_{B}T$ 
and $\sigma^{SC}=2.0\text{ }\mbox{\AA}$ in order to mimic various attractive
forces between the sidegroups.

Finally, a soft core (steric) repulsion, 
for all pairs involving $C-S$ and $N-S$ is added:
\begin{equation}
U_{repul}=\epsilon_{1}\Big(\frac{\sigma^{SC}}{r}\Big)^{12}.
\end{equation}
We should emphasise that the intra--molecular non--bonded interactions
are only included for pairs of sites connected via more than or equal
to three bonds.

\subsection{Potential energy function of the model $B$}

The second choice for the potential energy function 
outlined here provides a more phenomenological 
approach to the behaviour of the
coarse--grained systems from the basic geometrical principles.
For details of the Frenet--Serret picture of curves 
we refer the interested reader to Ref. \onlinecite{struik}.
We shall attempt to exploit and generalise Yamakawa's geometrical ideas for 
helical wormlike chain model \cite{yamakawa}.

Here the bond length potential and the non--bonded interactions 
retain the same functional form as in the model A.
The remaining bonded interactions are modeled as follows.

\subsubsection{Curvature}

The bond angle potential consists of a sum of harmonic terms involving 
the curvature $\kappa$:
\begin{equation}
U_{angle}=\sum_{angles}\frac{K_{\kappa}}{2}(\kappa-\kappa_{eq})^{2},
\end{equation}
where $K_{\kappa}=40.0\text{ } k_{B}T\cdot\mbox{\AA}^{-2}$ and 
$\kappa_{eq}=2.0\text{ }\mbox{\AA}$, $\kappa_{eq}=0.0\text{ }\mbox{\AA}$,
for angles centered in $C$ and $N$ respectively and 
with the curvature $\kappa$ defined as:
\begin{equation}
\kappa^{2} = 2\lbrack1-\cos(\theta)\rbrack.
\end{equation}
This definition of curvature is slightly different from the one 
used by Yamakawa as his definition also depends on
the pitch of the helix due to a different normalisation condition 
\cite{yamakawa}.

\subsubsection{Torsion}

Backbone is constrained towards a planar conformation by the terms:
\begin{equation}
U_{tors}=\sum_{dihedrals}\frac{K_{\tau}}{2}\tau^{2},
\end{equation}
involving only $C$ monomers, with $K_{\tau}=20.0\text{ } 
k_{B}T$.

\subsubsection{Chirality}

Chirality is introduced here via the term:
\begin{equation}
U_{chiral}=\frac{K_{\tau}}{2}(\tau-\tau_{0})^{2},
\end{equation}
applied to the quadruplets $S_{i}-C_i-C_{i+1}-S_{i+1}$ 
with $K_{\tau}=20.0\text{ } k_{B}T$.

\subsection{Procedure for Metropolis Monte Carlo simulations 
in continuous space}

For simulations we used the in--house 
Monte Carlo (MC) code named PolyPlus with the standard Metropolis 
algorithm \cite{metropolis} and local monomer moves, 
which represents a straightforward extension to a
more generic force (energy) field of the implementation 
described by us in Ref.~\onlinecite{CorFunc}. 
This was extensively used in the past and was quite successful
in tackling a wide range of problems for different heteropolymers
in solution. 

Note that the periodic boundary were unnecessary 
in the present study as we are dealing with an attractive
cluster. Therefore, no boundary conditions were required as 
the centre--of-mass of the system was maintained at the origin.

First, single coarse--grained $\beta$-strands made of $N=11$ residues are 
placed into a planar, antiparallel arrangement.
Starting from this initial conformation, systems of three different sizes 
(namely composed of $M=7,\ 15$ and $45$ strands) have been studied, 
using either the potential energy model $A$ or $B$ and 
with varied values of the chirality parameter $\tau_{0}$.
Specifically, we ran simulations in which $\tau_{0}$ takes values 
in the two sets $\{0.0,0.25,0.5,0.75,1.0\}$ and $\{0.0,0.1,0.2,0.3\}$
within the potential energy models $A$ and 
$B$ respectively. 
The difference between the two chosen sets is due to 
the different way in which the bonded interactions are 
implemented in both models.

Simulations times varied from $1\cdot10^{7}$ to $5\cdot10^{7}$ 
Monte Carlo sweeps.
About one fifth of that was required to achieve a good quality of equilibration, 
which was carefully monitored by analysing the trends in the potential energy,
the radius of gyration of the tape, and the wave number $k$ values;
and the rest four fifth of the run time were the production 
sweeps used for sampling of all observables. Thus, we were able to
achieve both a good equilibration and a good sampling statistics for the
observables of interest.


\section{\label{sec:defs}Definitions of some observables}

\begin{figure}
\centering
{\large\bf (a)} \hfill {\large\bf (b)}
\resizebox{15cm}{!}{
\includegraphics[angle=0, width = 1.0\textwidth]{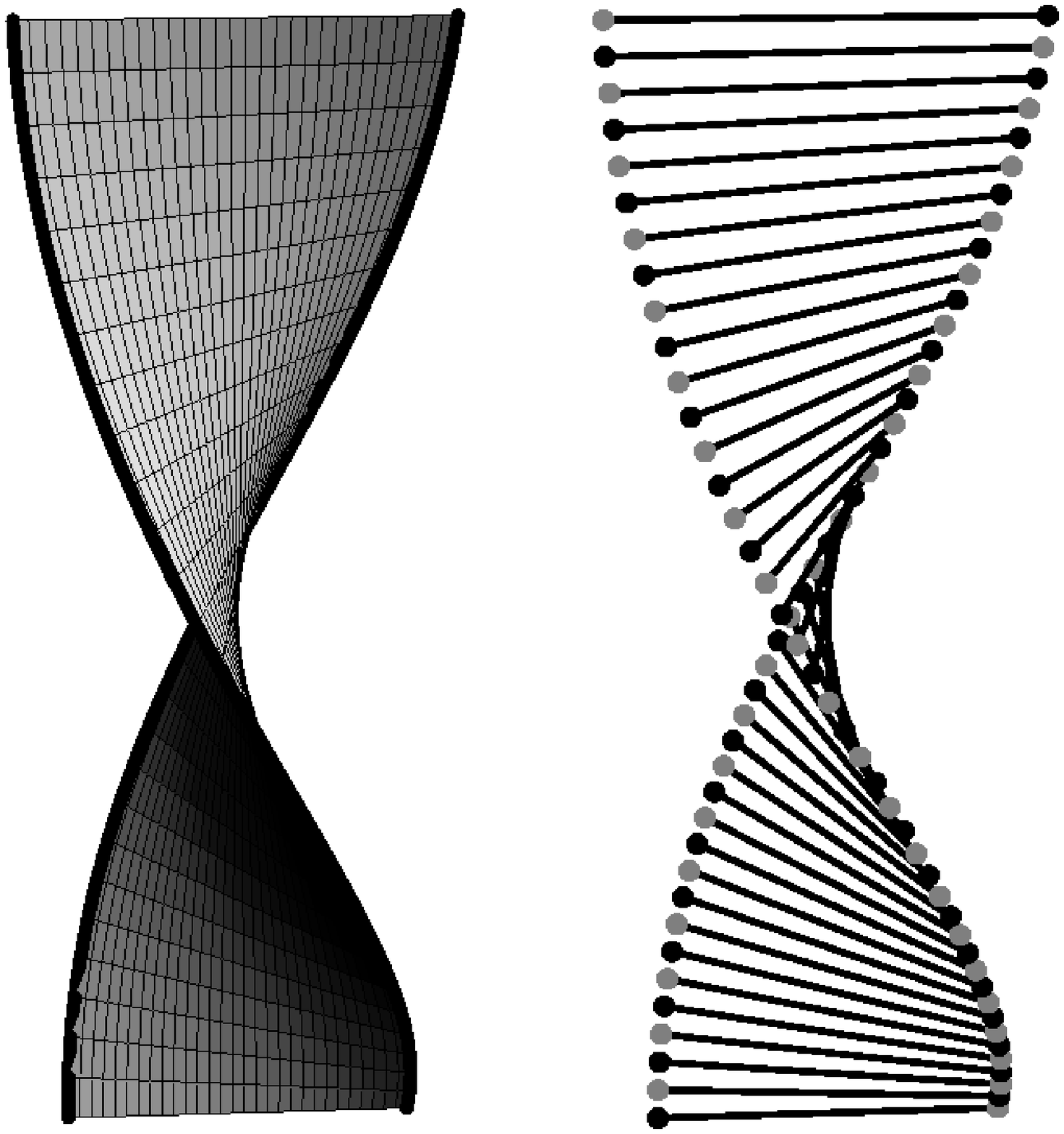}
}
\caption{
\label{fig:0}
}
\end{figure}

The circular helicoid is the minimal surface having a circular helix 
as its boundary \cite{struik}.
It can be described in the parametric form by:
\begin{equation}\label{eq:helicoid}
\begin{array}{ccl}
x & = & u\cos(kv), \\
y & = & u\sin(kv), \\
z & = & v.
\end{array}
\end{equation}
The circular helicoid (see Fig.~\ref{fig:0}a)
can be swept out by moving a segment in space, 
the length of this segment being equal to the length of the interval 
(domain) of the parameter $u$ definition.

The corresponding circular helix can be defined in a similar way
(thick lines in Fig.~\ref{fig:0}a):
\begin{equation}\label{eq:helix}
\begin{array}{ccl}
x & = & \rho\cos(kv), \\
y & = & \rho\sin(kv), \\
z & = & v.
\end{array}
\end{equation}
Here, the fixed radius $\rho$ is related to the parameter $u$, 
with $u \in [-\rho, \rho]$ and $k$ being defined as the {\it pitch wave number}
so that the {\it pitch} of a circular helix is $P=\frac{2\pi}{k}$.
The pitch wave number is, by convention, negative if the helix is 
left--handed and positive in the opposite case.

In order to find a connection between the helicoid and the tapes 
generated in our simulations, we require a consistent definition 
of the segment the motion of which in space sweeps the surface.
For this purpose, we can state that each strand's backbone 
lies along the vector
\begin{equation}
\bm{n}=[2\rho \cos(kz), 2\rho \sin(kz),z],
\end{equation}
where $\bm{z}$ is taken as the pitch axis of the tape.
Furthermore, we have to assume that $z$ varies in a discrete way 
along the tape with $\Delta z=d$, where $d$ is the {\it 
inter--strand distance}, (i.e. distance between the
nearest--neighbor strands). This somewhat modifies the calculation of 
the pitch, namely $P=\frac{2\pi d}{k}$.

Finally, three parameters are necessary to fully identify the vector 
$\bm{n}$ and the circular helix which delimits the surface.
The instantaneous values of $\rho$ (the {\it tape radius}), 
$k$ and $d$ are calculated by
taking the vector $\bm{x}_{ij}=(C_{2}^{i}-C_{10}^{j})$,
where $C_{l}^{i}$ is the position of monomer $C$ in the $l$th residue 
within the $i$th strand (see Fig.~\ref{fig:0}b for details).

The use of the vector $\bm{x}_{ii}$ is justified 
because the molecules behave themselves essentially as rigid rods.
In more details:
\begin{eqnarray}
\label{eq:rho}
\rho_{i} & = & \frac{\|\bm{x}_{ii}\|}{2}, \\
\label{eq:kappai}
k_{i,i+1} & = & \text{arccos}\Bigg(- \frac{\bm{x}_{ii}}
{\|\bm{x}_{ii}\|}\cdot \frac{\bm{x}_{i+1,i+1}}{\|\bm{x}_{i+1,i+1}\|}
\Bigg),\\
\label{eq:dd}
d_{i,i+1} & = & \frac{\|\bm{x}_{i,i+1} \cdot \bm{x}_{ii} 
\times \bm{x}_{i+1,i+1}\|}{\|\bm{x}_{ii} \times \bm{x}_{i+1,i+1}\|}
\end{eqnarray}
This calculation of the parameter $k$, which is related to a cosine, 
misses the correct evaluation of the sign.
To overcome this, an analogous measure related to a dihedral angle is needed.
The \emph{local dihedral angle} $(LDA)$ is defined as in Eq.~(\ref{eq:dihe}) 
with
\begin{eqnarray}
\label{eq:lda1}
\bm{r}_{12} & = & -\bm{x}_{ii}, \\
\label{eq:lda2}
\bm{r}_{32} & = & \bm{x}_{i,i+1}, \\
\label{eq:lda3}
\bm{r}_{34} & = & \bm{x}_{i+1,i+1}.
\end{eqnarray}
Monitoring the sign of this quantity gives information about 
the handedness of the helical cluster.

Thus, the complete formula for the calculation of the pitch wave number is:
\begin{equation}\label{eq:kappa}
k_{i,i+1} = \mbox{sign}(LDA)\cdot\text{arccos}\Bigg(- 
\frac{\bm{x}_{ii}}{\|\bm{x}_{ii}\|}\cdot 
\frac{\bm{x}_{i+1,i+1}}{\|\bm{x}_{i+1,i+1}\|}
\Bigg).
\end{equation}
In Tab.~\ref{tab:exp} we exhibit typical experimental values of the pitch 
wave number $k$ for some of oligopeptide--based supramolecular clusters.
These values were obtained from atomic force microscopy and 
transmission electron microscopy data.


\section{\label{sec:res}Results}

Tape structures in both models $A$ and $B$ appear
to be perfectly stable with single strands packed side by side along the 
backbone axis. Therefore, the simplistic representation of the 
hydrogen bonding adopted by us is successful in keeping the strands aligned.

With the increase of the chirality parameter $\tau_0 \neq 0$ 
the single strands acquire a somewhat regular twisted geometry.
The handedness and the magnitude of this twist have been 
quantified by calculating the value of the dihedral angles 
$\chi_{i}$ defined by the quadruplets 
$S_{i}-C_{i}-C_{i+2}-S_{i+2}$ where \cite{shamov} $i=2,3,4,...,8$.
Right--handed twist and left--handed twist are associated with 
positive and negative values of $\chi$ respectively.
The averaged values (over the production Monte
Carlo sweeps, over all but the terminal strands, and over the 
7 different $i$) of these 
dihedral values and the standard deviations are shown in Tabs.~\ref{tab:0a}
and \ref{tab:0b} for the models A and B respectively.
One can see a systematic increase of $\chi$ with the chirality
parameter $\tau_0$, irrespective of the model choice.

\begin{figure}
\centering
\resizebox{16cm}{!}{
\includegraphics[angle=-90, width = 1.0\textwidth]{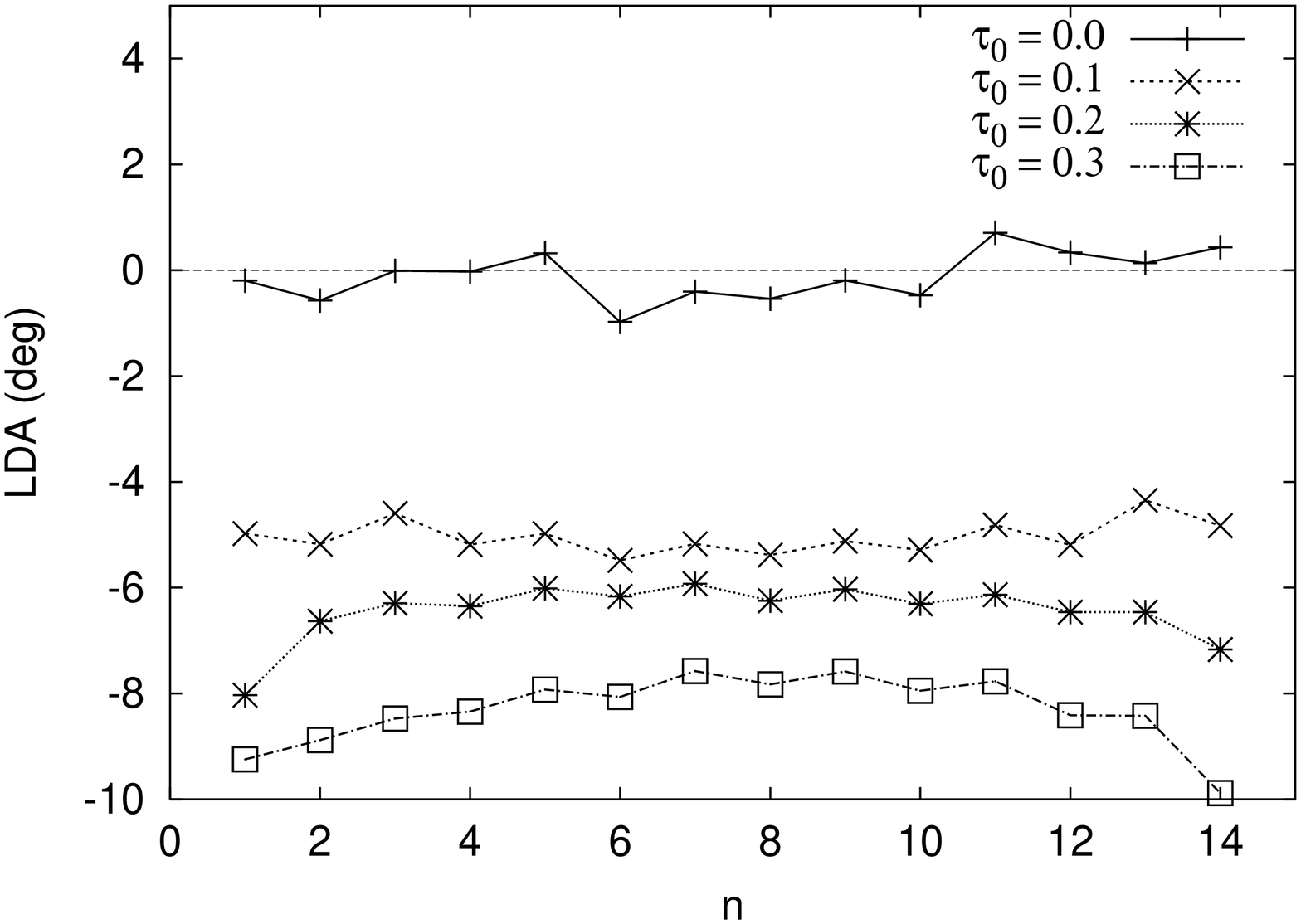}
}
\caption{
\label{fig:1}
}
\end{figure}

Increasing the chirality parameter $\tau_0$ leads also, as expected, 
to a persistent macroscopic twist of the tapes, which monotonically increases
with the value of $\tau_{0}$.
The numerical measure of the handedness of this twist 
could be expressed in terms of $LDA$ defined by Eqs.~(\ref{eq:dihe},%
\ref{eq:lda1},\ref{eq:lda2},\ref{eq:lda3}).
Fig.~\ref{fig:1} shows that, when chirality is introduced, the sign of LDA 
becomes well--defined and that the absolute value of LDA increases 
with $\tau_{0}$.
It is worthwhile to remark also, as a proof of the consistency of 
our procedure, that the achiral structure has no well--defined 
sign for $LDA$ (see Fig.~\ref{fig:1}).
Moreover, one can observe that after changing the sign 
of the chirality parameter $\tau_{0}$ the 
sign of $LDA$, and hence the handedness of the tape, will be
reversed (data not shown).

\begin{figure}
\centering
\resizebox{8cm}{!}{
\includegraphics[angle=-45, width = 0.4\textwidth, totalheight = 7cm]{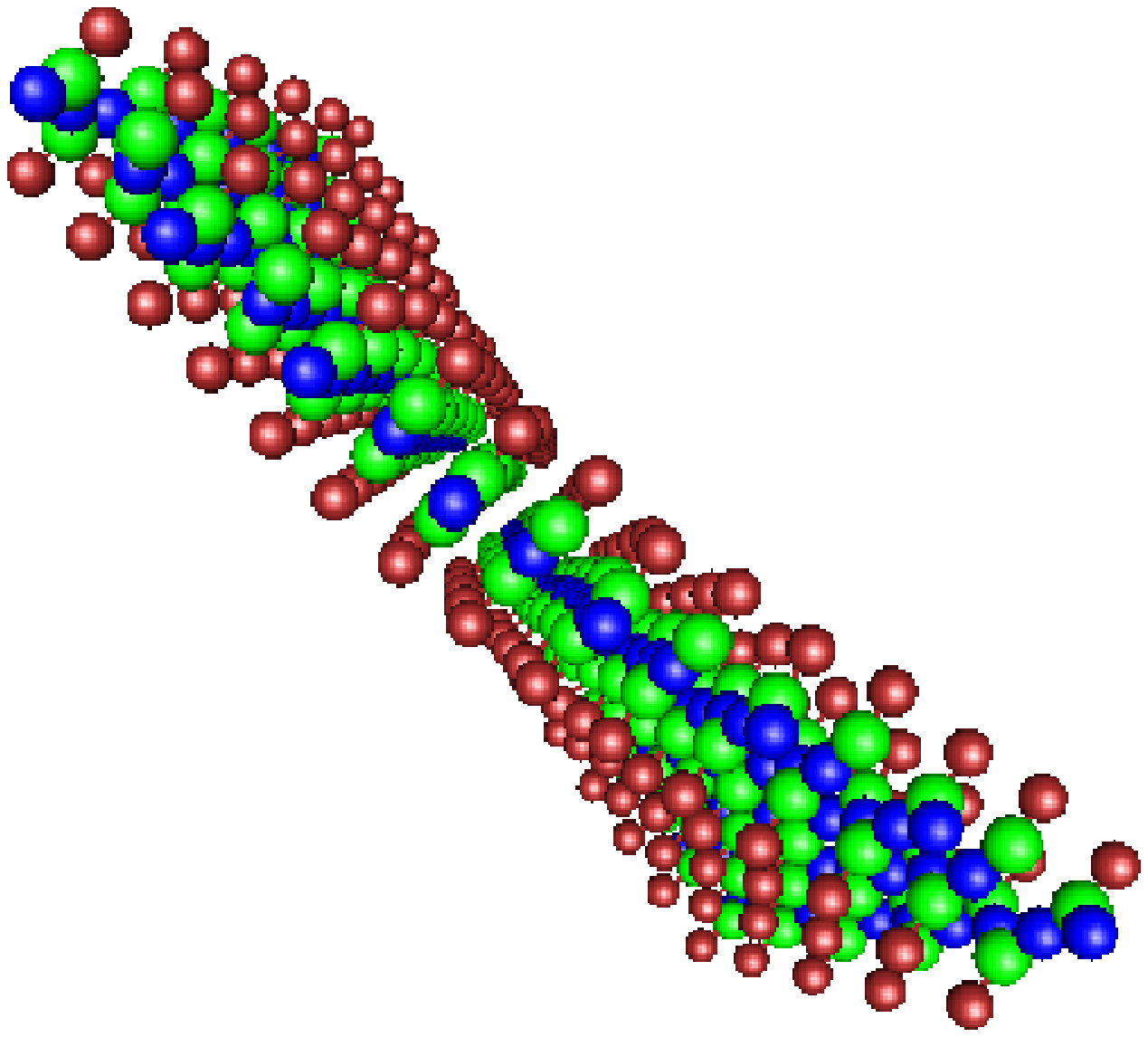}
}
\resizebox{8cm}{!}{
\includegraphics[angle=-45, width = 0.4\textwidth, totalheight = 7cm]{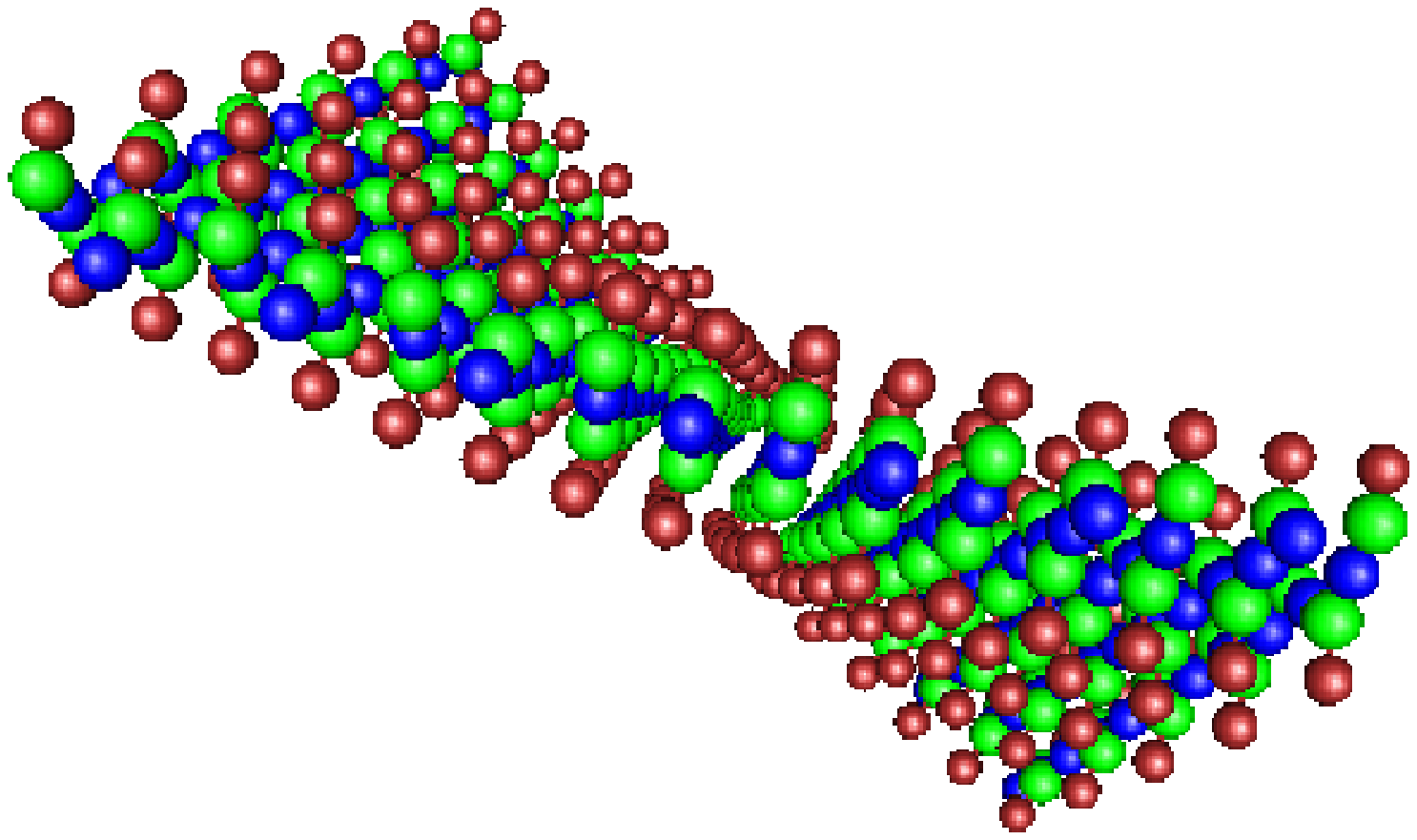}
}\\
\resizebox{7cm}{!}{
\includegraphics[angle=-180, width = 0.3\textwidth, totalheight = 5cm]{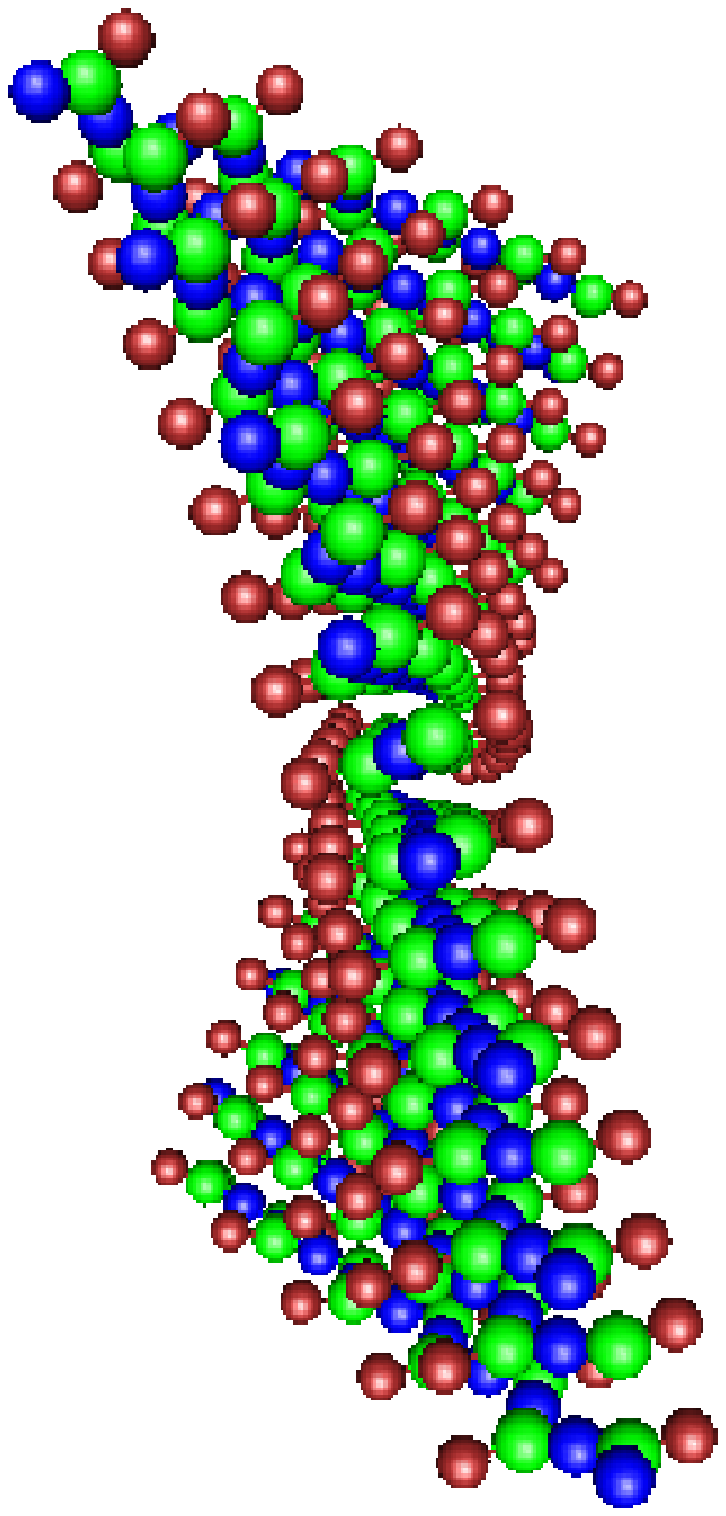}
}
\resizebox{7cm}{!}{
\includegraphics[angle=-180, width = 0.3\textwidth, totalheight = 5cm]{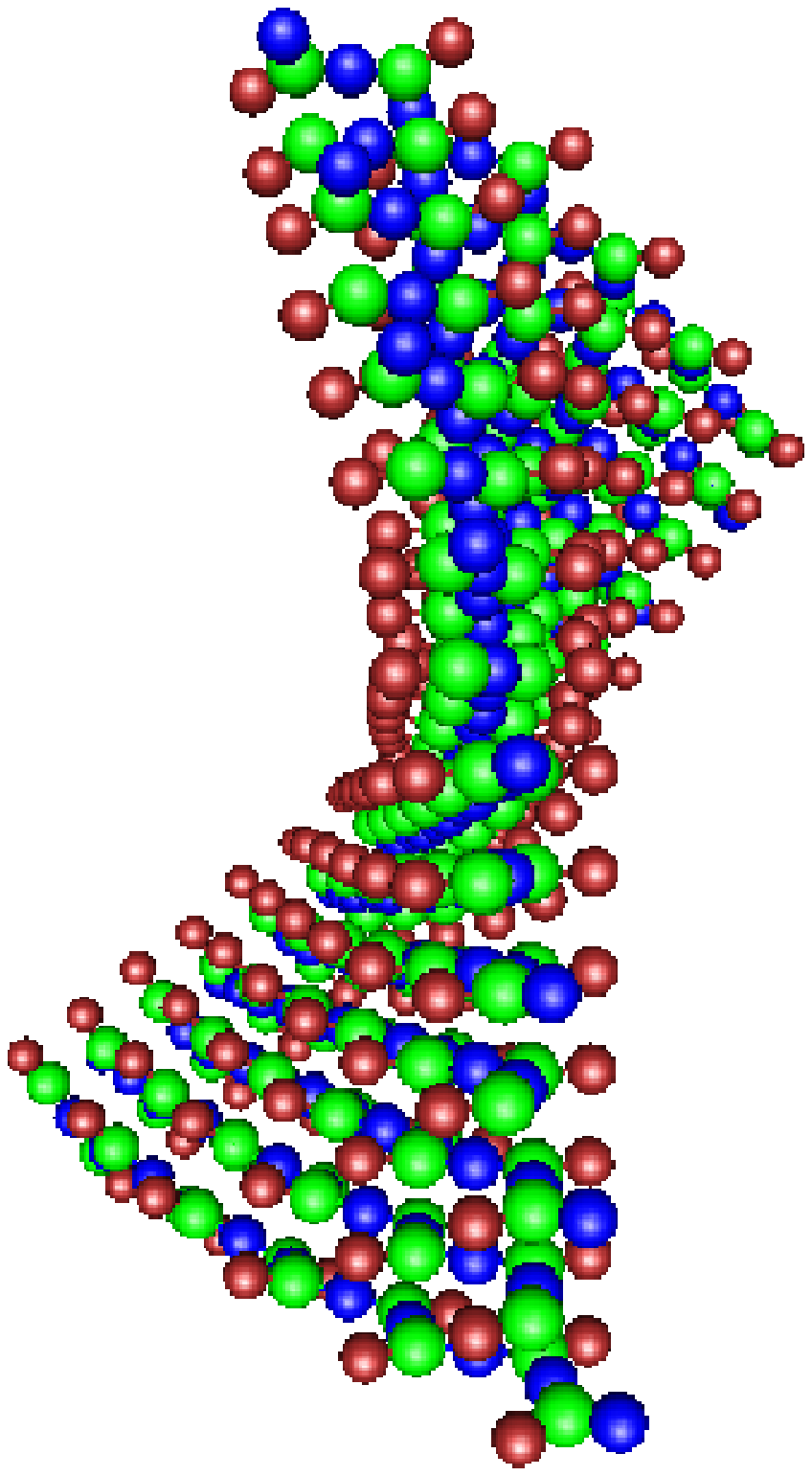}
}
\caption{
\label{fig:2}
}
\end{figure}

\begin{figure}
\centering
\resizebox{8cm}{!}{
\includegraphics[angle=-90, width = 0.5\textwidth, totalheight = 8cm]{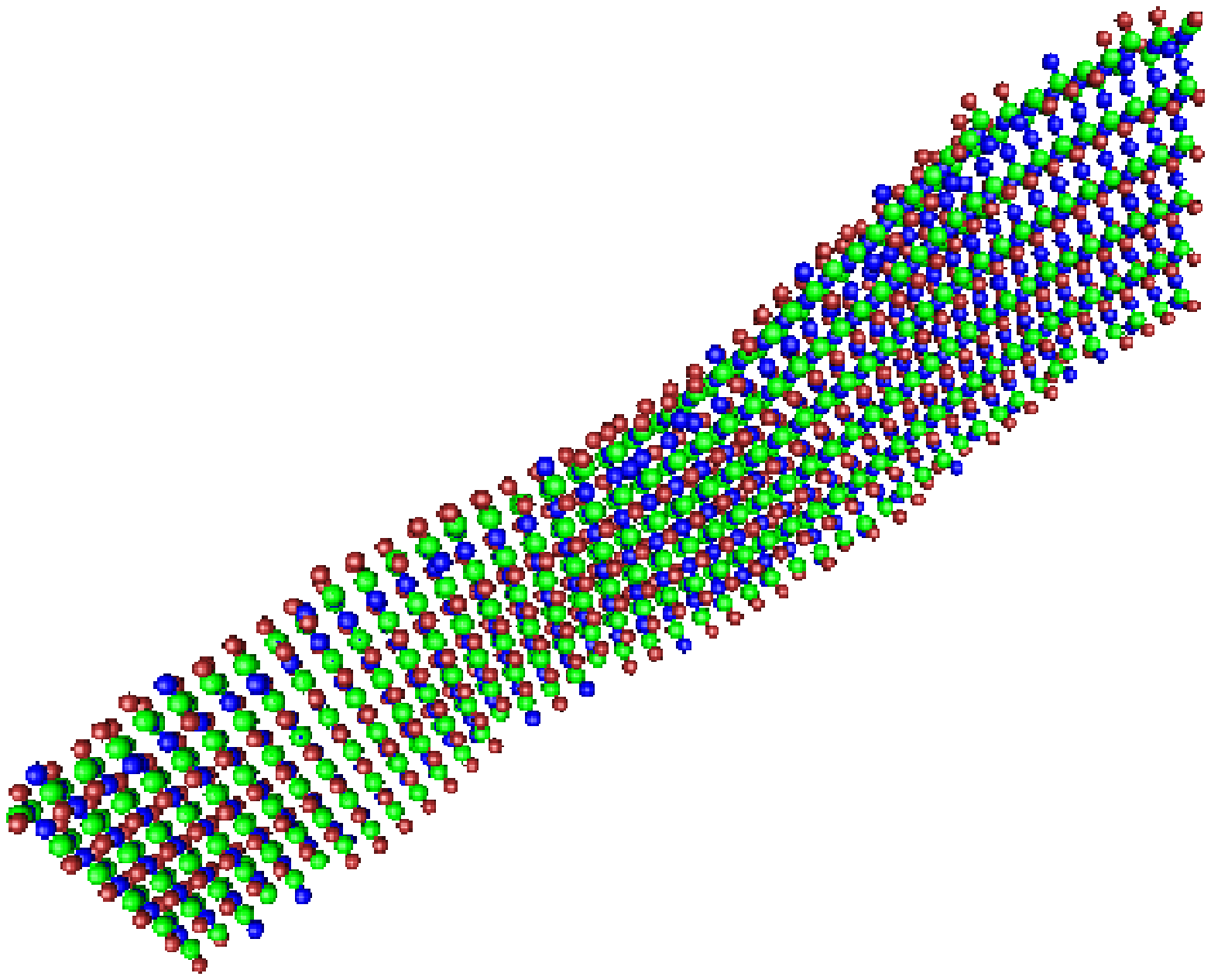}
}\\
\resizebox{8cm}{!}{
\includegraphics[angle=-90, width = 0.5\textwidth, totalheight = 8cm]{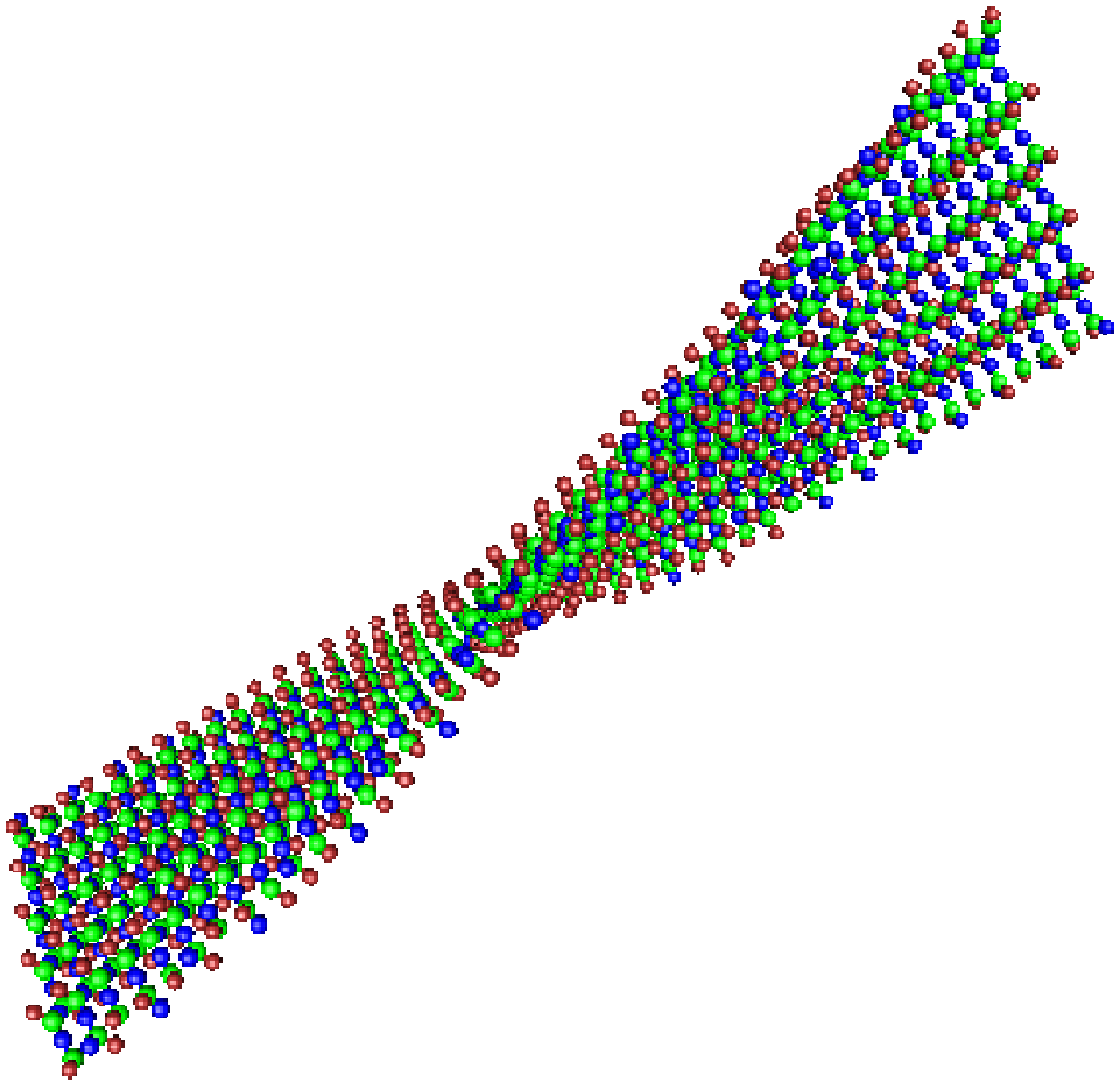}
}\\
\resizebox{8cm}{!}{
\includegraphics[angle=-90, width = 0.5\textwidth, totalheight = 8cm]{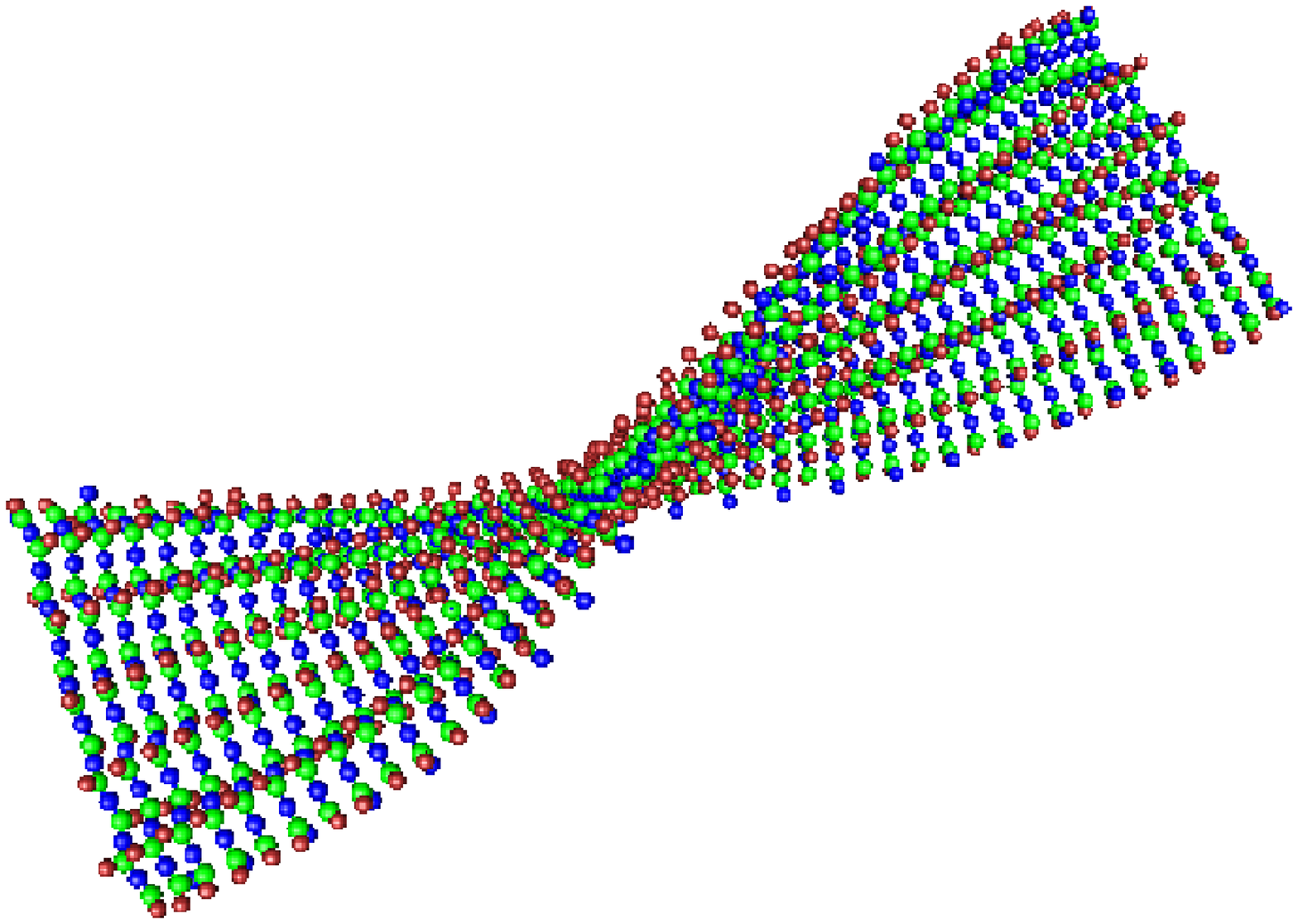}
}
\caption{
\label{fig:3}
}
\end{figure}

Figs.~\ref{fig:2},\ref{fig:3} also show averaged
equilibrium snapshots related 
to the systems with different values of the chirality parameter $\tau_{0}$
in the models $A$ and  $B$ respectively. 
>From these one could see how the structures change from a flat into more
and more twisted tapes as $\tau_{0}$ increases.

Next, we would like to compare our simulated structures with 
the geometry of a left--handed circular helicoid, the definition for which was 
given in Sec.~\ref{sec:defs}. Specifically, we are 
interested in characterizing the circular helices which sweep
the boundaries of that surface.
The details of calculation of the three parameters, $k$, $d$ and $\rho$,
which are necessary for connection of the \emph{idealized} 
geometry with that of the simulated tapes, can be also found in 
Sec.~\ref{sec:defs} and in Fig.~\ref{fig:0}.
These values we can calculate from the coordinates 
of each two consequent strands.

Clearly, the chains at the boundaries of the tape behave 
in a somewhat different way from those buried inside. 
More generally, despite of the intra--molecular origin of chirality, 
the conformation of single strands within a tape also depends on their
interactions with nearest--neighboring strands. As chirality is increased,
the deformation of the flat geometry of a single strand progresses.
Therefore, for fairly short tapes with small $M$ we could quite
significant finite size effects leading to considerable deviations from
the regular geometrical surfaces.

Thus, we shall compute the average values and the standard deviations
of the quantities  $k$, $d$ and $\rho$ over the span of the tape
(with the exception of the two terminal strands on both edges
to reduce the boundary effects), as well as over the production sweeps,
in order to understand at what extent they vary along the tape.
These values are presented in Tabs.~\ref{tab:1} and \ref{tab:2}
for the models $A$ and $B$ respectively.

The values of the helix radius $\rho$ does not seem to vary significantly
along the tape, which is reflected in a relatively small value
of its standard deviation $\sigma_{\rho}$. Evidently, the average
value of $\rho$ is essentially independent of the number of strands
$M$ or the chirality parameter $\tau_0$ as it is related to the 
conformation of a single strand. While the standard deviation
of the distance between two strands $d$ is relatively large, we
do not observe any systematic dependencies of its values on either
the location within the tape or the value of the chirality 
parameter $\tau_0$. A large value of $\sigma_d$ could be 
attributed to the inter--molecular interactions contribution
to the distances between nearest close--packed strands.

\begin{figure}
\centering
\resizebox{15cm}{!}{
\includegraphics[angle=-90, width = 0.5\textwidth, totalheight = 4.4cm]{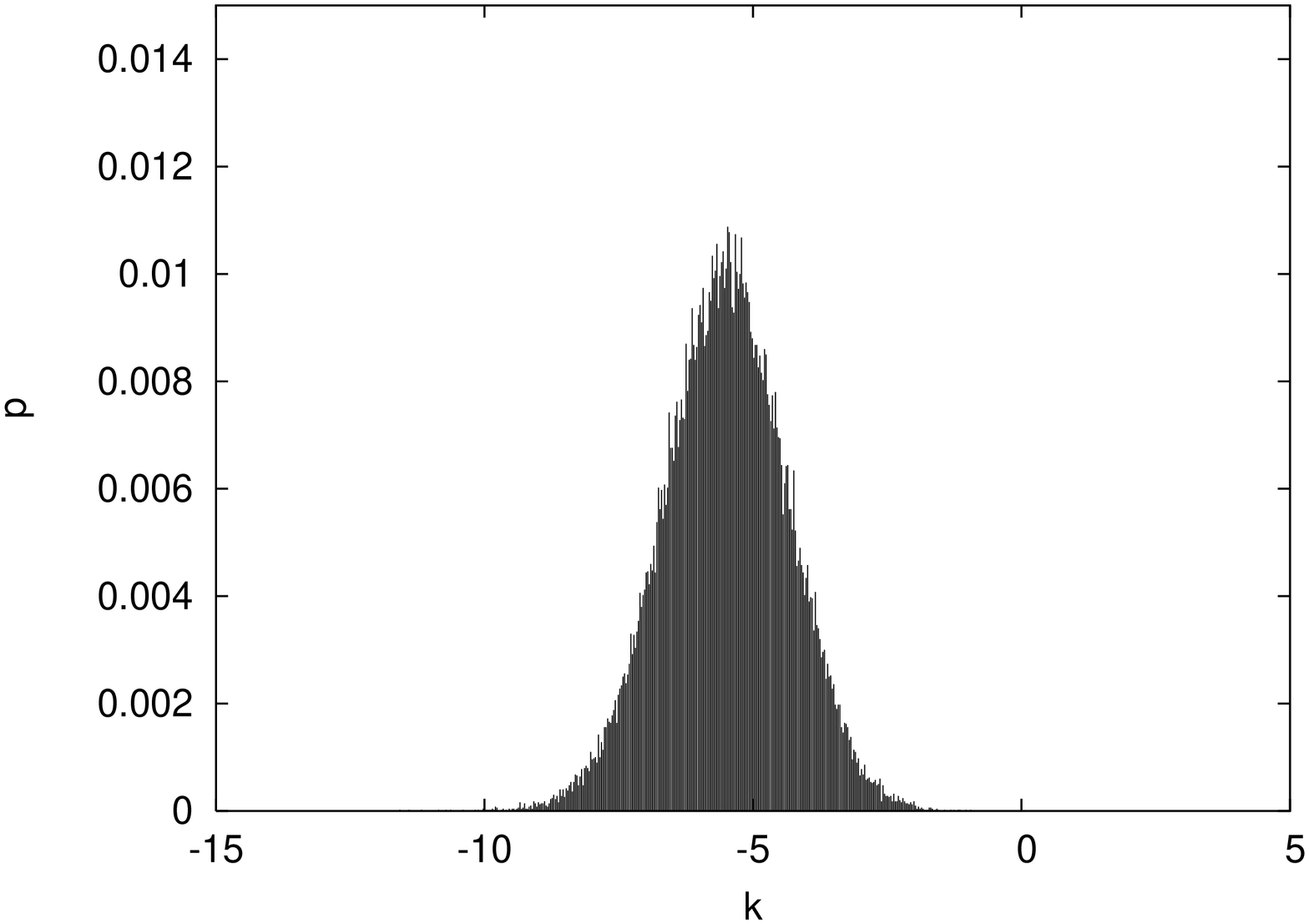}
}
\resizebox{15cm}{!}{
\includegraphics[angle=-90, width = 0.5\textwidth, totalheight = 4.4cm]{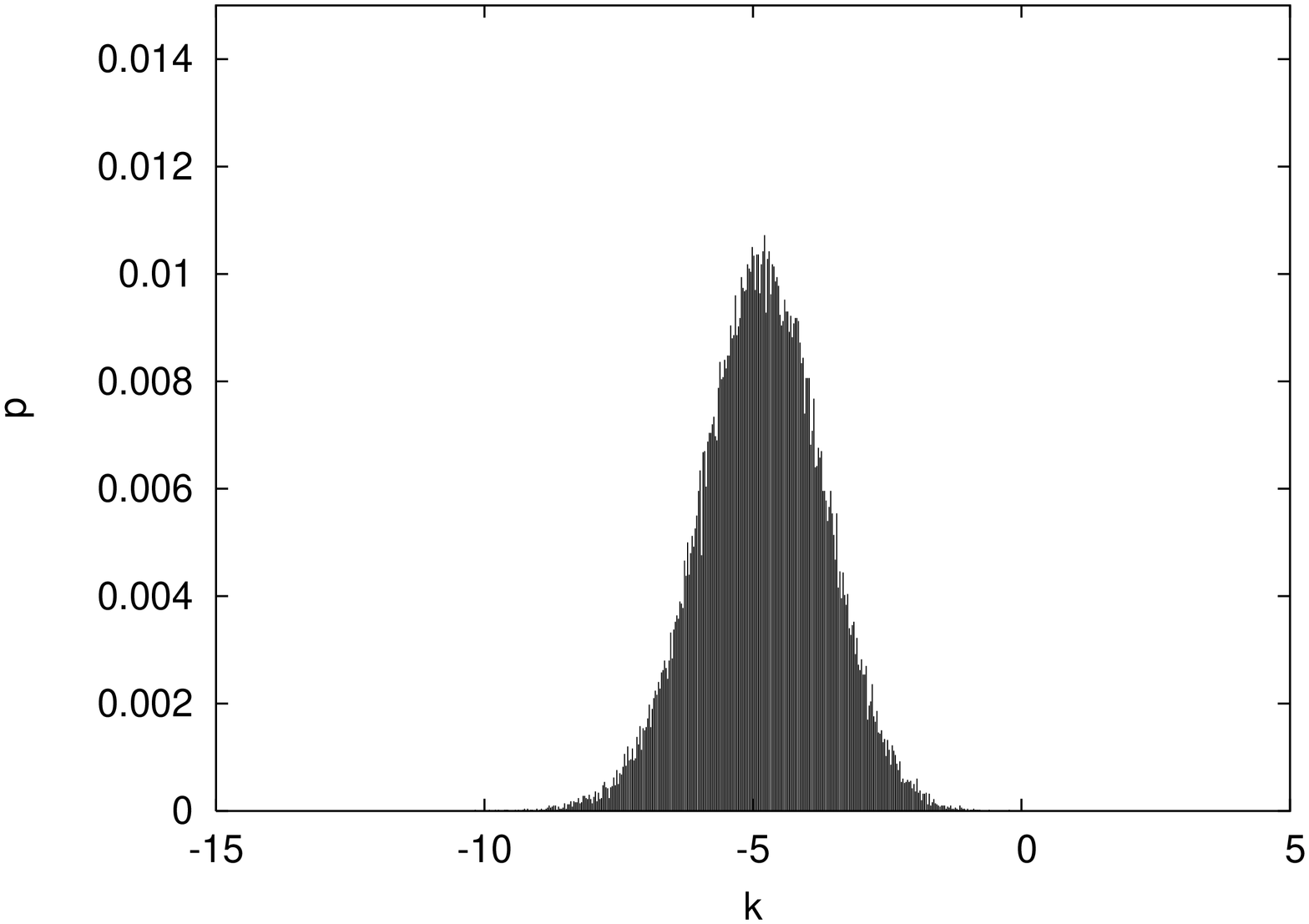}
}
\resizebox{15cm}{!}{
\includegraphics[angle=-90, width = 0.5\textwidth, totalheight = 4.4cm]{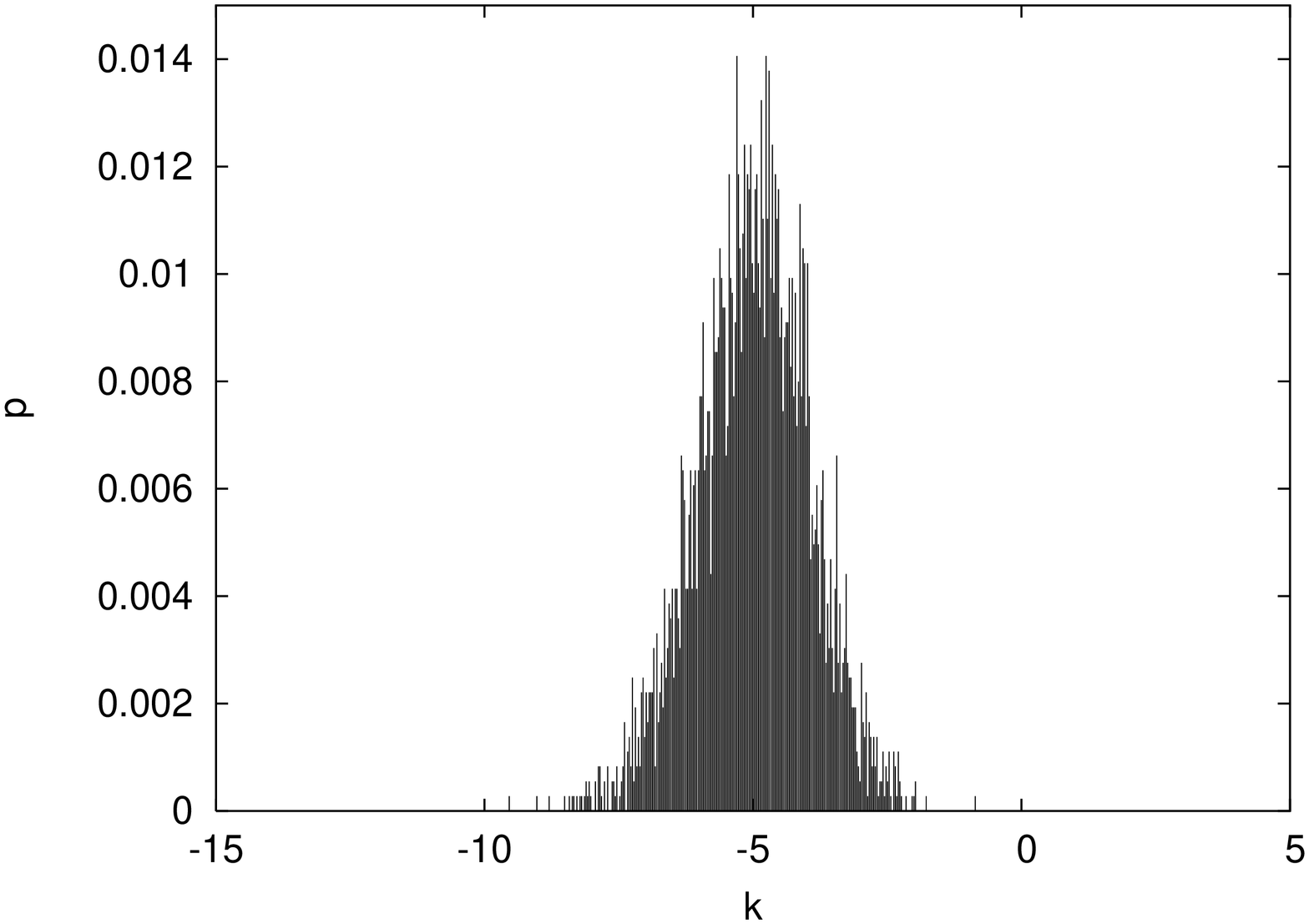}
}
\caption{
\label{fig:4}
}
\end{figure}

However, the pitch wave number $k$ is strongly dependent on the
location of the strands pair used in its calculation inside the tape,
which is especially striking for small systems made of $M=7$ and $M=15$ 
chains since they do not as yet complete a full turn of the helicoid. 
The results of the calculations of the average and 
standard deviation of the pitch wave number $k$ shown in Tab.~\ref{tab:1},
Tab.~\ref{tab:2} were thus obtained by taking only the three 
central strands, which has the advantage that the results become 
less sensitive to the edge--effects.
Clearly, the central area of the tape of different sizes $M$ behaves, 
as far as the pitch wave number is concerned, in a similar way
and numerically approaches the value of $k$ in the tape of $M=45$ strands.
This can also be seen from the histograms (probability
distributions) of $k$ in Fig.~\ref{fig:4}. Note that the location of
the peak of these shifts to the right with $M$ somewhat. 
The values of $k$ which we have obtained in the range of the
studied chirality parameter $\tau_0$ choices correspond to the experimental
values of $k$ shown in Tab. \ref{tab:exp}.
Thus, we need about $3-5\ k_B T$ for $K_{\tau}$ in our model
in order to obtain the highest of the known experimental values of $k$.

To check the quality of our parameters, we then performed, 
for each system, a self--consistent fitting procedure, in which we 
considered two data sets, $r_1$ and $r_2$, 
which are related to the two respective edges 
of the tape. These data sets comprised a sequence of
the average positions \cite{footnote1}.
of the penultimate monomers in consecutive strands within the 
tape. Note that the penultimate monomers were considered rather 
than the end  monomers to reduce the ``end effects''. Also, because 
the strands were assembled in an antiparallel pattern, one typical 
sequence of positions would comprise monomers 
$C_{2}^{1}$, $C_{10}^{2}$, $C_{2}^{3}$, $C_{10}^{4}$, \dots, $C_{2}^{M-1}$. 
These were fitted with a regular helix described by Eq.~(\ref{eq:helix})
sweeping the end of the regular helicoid with the
parameters $d$, $\rho$ and $k$ calculated from the simulation data.
  
The fitting procedure for each system produces, as the final output, 
the two mean displacements $\langle\Delta\rangle_{r_1}$ and $\langle\Delta
\rangle_{r_2}$ 
between the \emph{regular} helix and the data sets $r1$ and $r2$ respectively.
Since the data sets $r_1$ and $r_2$ are statistically equivalent 
we calculated the mean displacement of our points from the regular
helix, $\langle\Delta\rangle$, averaged over the two values.
As can be seen from Tab.~\ref{tab:3}, the resulting mean displacements
are relatively small compared to the size of the van der Waals radius
of various monomers for the systems made of $M=7$ and $M=15$.
Therefore, our overall procedure is quite satisfactory.
Note that a somewhat larger values of $\langle\Delta\rangle$ for the
systems made of  $M=45$
strands can be related to the need for a better equilibration in this
largest of the studied systems, which was also seen from 
monitoring the trends in the global observables such as the
squared radius of gyration of the tape.

Thus, overall, we conclude 
that both of the potential models suggested here are successful in generating
chirality within a stable tape cluster.



\section{Conclusion}

In this paper we have proposed two coarse--grained models for 
short peptides in an extended $\beta$--strand--like conformation.
We also have studied these strands self--assembled into
a supramolecular $\beta$-tape in case of the model oligopeptides
with identical side groups attached.
A fine tuned combination of Lennard--Jones potential terms was 
successful in stabilizing the chains within such a 
two--dimensional structure.

Chirality was then introduced on a molecular level with resulting in 
a regular twist of the surface of the tape.
Within the two different models we have investigated
the effect of changing the values of the chirality parameter $\tau_{0}$.
As we only considered homogeneous
sequences yielding tapes with identical sides,
the equilibrium structures obtained at the end of the simulations 
had a geometry of a circular helicoid with the pitch wave number
$k$ increasing linearly with $\tau_{0}$.

In model $A$ the chirality term is added as a simple asymmetric 
contribution to the three--folded dihedral potential which is typically used
in coarse--grained models of $\beta$-sheets \cite{thiru1,thiru2}. 
The remaining bonded interactions in their analytical expressions 
and in the numerical strength were taken akin to those of the
fully atomistic force fields. Therefore, in essence, here we were introducing
chirality into a well--established potential energy model.

Model $B$, conversely, is more coarse--grained and relies on the
principles of the differential geometry of curves and surfaces 
\cite{harris}. Importantly, in this model we still have obtained the
results comparable to those of the more detailed model $A$.
This establishes a degree of universality in the transfer of chirality 
from the intra--molecular to the supramolecular level.
Despite the difference in the way how chirality was introduced 
in the both models, a macroscopic regular twist was generated
equivalently.

Both models could be easily extended to include hydrophobic/hydrophilic
and explicitly charged sidegroups leading to the difference of the tapes
sides, something we would like to study in the future.
Such an extended study of different coarse--grained
oligopeptide sequences of interest should allow us to describe
higher order self--assembled structures (ribbons, fibers and fibrils)
in detail, providing valuable insights for the experiment.

\begin{acknowledgments}

The authors would like to thank Professor Neville Boden,
Professor Alexei Kornyshev,
Professor Alexander Semenov, Dr Amalia Aggeli, Dr Colin Fishwick,  
as well as our colleagues
Dr Ronan Connolly and Mr Nikolaj Georgi for useful discussions.
Support from the IRCSET basic research grant SC/02/226
is also gratefully acknowledged.

\end{acknowledgments}


\newpage

\newpage

\begin{table}[!hbp]
\begin{center}
\begin{tabular}{|c|c|c|c|c|}
\hline
Peptide name & Primary Structure & $k(\text{deg})$ & Experimental Technique  & Reference \\
\hline
P$_{11}-$I & {\small QQRQQQQQEQQ} &  $-3.0$ & {\small TEM} & \cite{soms2} \\
P$_{11}-$II & {\small QQRFQWQFEQQ} & $-1.0$ & {\small TEM} & \cite{soms2} \\
KFE8 & {\small FKFEFKFE} & $-8.7$ & {\small AFM} & \cite{mit1,mit2} \\
A$\beta(10-35)$ & {\small YEVHHQKLVFFAEDVGSNKSAIIGLM} & $-1.6$ & {\small TEM} & \cite{amyl} \\
\hline
\end{tabular}
\caption{\label{tab:exp}
Values of the pitch wave number $k$ obtained from the experimental analysis on 
chiral supramolecular clusters formed from several synthetic and natural 
peptides.
}
\end{center}
\end{table}
\newpage

\begin{table}[!hbp]
\begin{center}
\begin{tabular}{|c|c|c|}
\hline
\multicolumn{3}{|c|}{$M=7$} \\
\hline
$\tau_{0}$ & $\chi(\text{deg})$ & $\sigma_{\chi}$\\
\hline
$0.00$ &  $0.11$ & $0.04$ \\
$0.25$ &  $3.77$ & $0.29$ \\
$0.50$ &  $7.34$ & $0.50$ \\
$0.75$ & $10.50$ & $0.63$ \\
$1.00$ & $13.45$ & $0.75$ \\
\hline
\multicolumn{3}{|c|}{$M=15$} \\
\hline
$\tau_{0}$ & $\chi(\text{deg})$ & $\sigma_{\chi}$\\
\hline
$0.00$ &  $0.05$ & $0.02$ \\
$0.25$ &  $3.79$ & $0.27$ \\
$0.50$ &  $6.90$ & $0.41$ \\
$0.75$ &  $9.52$ & $0.53$ \\
$1.00$ & $12.01$ & $0.67$ \\
\hline
\multicolumn{3}{|c|}{$M=45$} \\
\hline
$\tau_{0}$ & $\chi(\text{deg})$ & $\sigma_{\chi}$\\
\hline
$0.00$ &  $0.19$ & $0.01$ \\
$0.25$ &  $4.39$ & $0.28$ \\
$0.50$ &  $6.95$ & $0.43$ \\
$0.75$ &  $8.42$ & $0.50$ \\
$1.00$ & $11.68$ & $0.74$ \\
\hline
\end{tabular}
\caption{\label{tab:0a}
Average value of the individual strand chirality angle $\chi$
and its standard deviation $\sigma_{\chi}$ 
obtained from Monte Carlo simulations in the
potential model $A$.}
\end{center}
\end{table}

\newpage

\begin{table}[!hbp]
\begin{center}
\begin{tabular}{|c|c|c|}
\hline
\multicolumn{3}{|c|}{$M=7$} \\
\hline
$\tau_{0}$ & $\chi(\text{deg})$ & $\sigma_{\chi}$\\
\hline
$0.0$ &   $0.08$ & $0.05$ \\
$0.1$ &   $7.16$ & $0.67$ \\
$0.2$ &  $13.22$ & $1.51$ \\
$0.3$ &  $18.73$ & $2.22$ \\
\hline
\multicolumn{3}{|c|}{$M=15$} \\
\hline
$\tau_{0}$ & $\chi(\text{deg})$ & $\sigma_{\chi}$\\
\hline
$0.0$ &   $0.09$ & $0.02$ \\
$0.1$ &   $6.85$ & $0.66$ \\
$0.2$ &  $12.39$ & $1.44$ \\
$0.3$ &  $17.30$ & $2.35$ \\
\hline
\multicolumn{3}{|c|}{$M=45$} \\
\hline
$\tau_{0}$ & $\chi(\text{deg})$ & $\sigma_{\chi}$\\
\hline
$0.0$ &   $0.42$ & $0.07$ \\
$0.1$ &   $6.20$ & $0.74$ \\
$0.2$ &  $12.21$ & $1.62$ \\
$0.3$ &  $16.73$ & $2.47$ \\
\hline
\end{tabular}
\caption{\label{tab:0b}
Average value of the individual strand chirality angle $\chi$
and its standard deviation $\sigma_{\chi}$ 
obtained from Monte Carlo simulations in the
potential model $B$.}
\end{center}
\end{table}

\newpage

\begin{table}[!hbp]
\begin{center}
\begin{tabular}{|c|c|c|c|c|c|c|c|}
\hline
\multicolumn{7}{|c|}{$M= 7$} \\
\hline
$\tau_{0}$ & $k(\text{deg})$ & $\sigma_{k}$ & $d(\mbox{\AA})$ 
& $\sigma_{d}$ & $\rho(\mbox{\AA})$ & $\sigma_{\rho}$\\
\hline
$0.25$ & $-2.2$ & $1.0$ & $1.97$ & $0.35$ & $13.6$ & $0.2$ \\
$0.5$ & $-3.3$ & $1.2$ & $2.00$ & $0.31$ & $13.5$ & $0.2$ \\
$0.75$ & $-4.5$ & $1.2$ & $2.10$ & $0.30$ & $13.5$ & $0.2$ \\
$1.0$ & $-5.5$ & $1.2$ & $2.01$ & $0.29$ & $13.4$ & $0.2$ \\
\hline
\multicolumn{7}{|c|}{$M= 15$} \\
\hline
$\tau_{0}$ & $k(\text{deg})$ & $\sigma_{k}$ & $d(\mbox{\AA})$ 
& $\sigma_{d}$ & $\rho(\mbox{\AA})$ & $\sigma_{\rho}$ \\
\hline
$0.25$ & $-2.2$ & $0.9$ & $2.06$ & $0.37$ & $13.6$ & $0.2$ \\
$0.5$ & $-3.1$ & $1.1$ & $1.81$ & $0.22$ & $13.6$ & $0.2$ \\
$0.75$ & $-3.9$ & $1.2$ & $2.20$ & $0.30$ & $13.6$ & $0.2$ \\
$1.0$ & $-4.8$ & $1.2$ & $2.01$ & $0.30$ & $13.5$ & $0.2$ \\
\hline
\multicolumn{7}{|c|}{$M=45$} \\
\hline
$\tau_{0}$ & $k(\text{deg})$ & $\sigma_{k}$ & $d(\mbox{\AA})$ 
& $\sigma_{d}$ & $\rho(\mbox{\AA})$ & $\sigma_{\rho}$\\
\hline
$0.25$ & $-2.2$ & $1.2$ & $2.35$ & $0.40$ & $13.6$ & $0.2$ \\
$0.5$ & $-3.2$ & $1.2$ & $2.44$ & $0.34$ & $13.6$ & $0.2$ \\
$0.75$ & $-4.0$ & $1.0$ & $2.43$ & $0.31$ & $13.6$ & $0.2$ \\
$1.0$ & $-5.0$ & $1.0$ & $2.39$ & $0.26$ & $13.5$ & $0.2$ \\
\hline
\end{tabular}
\caption{\label{tab:1}
Average value and standard deviation $\sigma$, 
obtained from Monte Carlo simulations, for the helical parameters $k$ 
(Eqs. \ref{eq:kappai} and \ref{eq:kappa}), $d$ (Eqs. \ref{eq:dd}) 
and $\rho$ (Eqs. \ref{eq:rho}) for the potential energy model $A$.}
\end{center}
\end{table}
\newpage
\begin{table}[!hbp]
\begin{center}
\begin{tabular}{|c|c|c|c|c|c|c|}
\hline
\multicolumn{7}{|c|}{$M= 7$} \\
\hline
$\tau_{0}$ & $k(\text{deg})$ & $\sigma_{k}$ & $d(\mbox{\AA})$ 
& $\sigma_{d}$ & $\rho(\mbox{\AA})$ & $\sigma_{\rho}$ \\
\hline
$0.1$ & $-2.6$ & $1.2$ & $1.86$ & $0.31$ & $13.7$ & $0.2$ \\
$0.2$ & $-3.8$ & $1.3$ & $2.00$ & $0.29$ & $13.7$ & $0.2$ \\
$0.3$ & $-5.0$ & $1.2$ & $1.95$ & $0.25$ & $13.6$ & $0.2$ \\
\hline
\multicolumn{7}{|c|}{$M= 15$} \\
\hline
$\tau_{0}$ & $k(\text{deg})$ & $\sigma_{k}$ & $d(\mbox{\AA})$ & 
$\sigma_{d}$ & $\rho(\mbox{\AA})$ & $\sigma_{\rho}$ \\
\hline
$0.1$ & $-2.5$ & $1.1$ & $2.12$ & $0.33$ & $13.7$ & $0.2$ \\
$0.2$ & $-3.4$ & $1.2$ & $2.15$ & $0.31$ & $13.7$ & $0.2$ \\
$0.3$ & $-4.2$ & $1.2$ & $2.19$ & $0.26$ & $13.7$ & $0.2$ \\
\hline
\multicolumn{7}{|c|}{$M= 45$} \\
\hline
$\tau_{0}$ & $k(\text{deg})$ & $\sigma_{k}$ & $d(\mbox{\AA})$ 
& $\sigma_{d}$ & $\rho(\mbox{\AA})$ & $\sigma_{\rho}$ \\
\hline
$0.1$ & $-2.4$ & $1.1$ & $2.25$ & $0.26$ & $13.7$ & $0.2$\\
$0.2$ & $-3.6$ & $1.2$ & $2.46$ & $0.31$ & $13.7$ & $0.2$\\
$0.3$ & $-4.2$ & $1.2$ & $2.47$ & $0.26$ & $13.7$ & $0.2$\\
\hline
\end{tabular}
\caption{\label{tab:2}
Average value and standard deviation $\sigma$, 
obtained from Monte Carlo simulations, for the helical parameters $k$
(Eqs. \ref{eq:kappai} and \ref{eq:kappa}), $d$ (Eqs. \ref{eq:dd}) 
and $\rho$ (Eqs. \ref{eq:rho}) 
for the potential energy model $B$.}
\end{center}
\end{table}
\newpage
\begin{table}[!hbp]
\begin{center}
\begin{tabular}{|c|c|c|c|}
\hline
\multicolumn{4}{|c|}{$M= 7$} \\
\hline
$\tau_{0}$ & $0.1$ & $0.2$ & $0.3$ \\
\hline
$\langle\Delta\rangle (\mbox{\AA})$ & $0.9$ & $0.65$ & $0.55$\\
\hline
\multicolumn{4}{|c|}{$M= 15$} \\
\hline
$\tau_{0}$ & $0.1$ & $0.2$ & $0.3$ \\
\hline
$\langle\Delta\rangle (\mbox{\AA})$ & $0.65$ & $0.9$ & $0.3$\\
\hline
\multicolumn{4}{|c|}{$M= 45$} \\
\hline
$\tau_{0}$ & $0.1$ & $0.2$ & $0.3$ \\
\hline
$\langle\Delta\rangle (\mbox{\AA})$ & $1.0$ & $1.4$ & $1.2$\\
\hline
\end{tabular}
\caption{\label{tab:3}
Fitting results for potential energy model $B$. 
$\langle\Delta\rangle$ is the mean displacement between the regular geometrical
and simulated helices structures. 
A similar behaviour for the potential energy model $A$ has been found also.}
\end{center}
\end{table}

$\left.\right.\left.\right.\left.\right.\left.\right.\left.\right.\left.\right.$
\newpage

\centerline{{\Large\bf Figure Captions}}

\noindent{\large\bf Fig. 1. \ }
\begin{quotation}\noindent
(a) A circular helicoid described in parametric form by Eq. 
(\ref{eq:helicoid}).
The constants used to generate the surface were obtained 
from Monte Carlo simulations of the system of size $M=45$ with
and chirality parameter $\tau_0=0.3$ using the potential energy model $B$.
The thick lines (helical curves) sweeping the two surface's edges are 
described in parametric form by Eq. (\ref{eq:helix}).
\\
(b)A schematic representation of the regular tape 
corresponding to the circular helicoid.
Gray and black points represent the positions of the monomers 
$C^{i}_{2}$ and $C^{i}_{10}$ $(\text{with }i=2,M-1)$ respectively.
The connecting lines correspond to the vectors 
$\bm{n}_{i}=[2\rho \cos(kv), 2\rho \sin(kv),v\,d]$ $(\text{with }i=2,M-1)$,
where the values for $\rho\text{, }k\text{ and }d$ are, once again, 
taken from Monte Carlo simulations
and $v=2,3,4,...,M-1$.
The positions of the monomers $C^{i}_{2}$ and $C^{i}_{10}$ were used 
as the reference data in our fitting procedure.
\end{quotation}

\noindent{\large\bf Fig. 2. \ }
\begin{quotation}\noindent
Plot of the average LDA (Eqs. \ref{eq:dihe} and \ref{eq:lda1}, 
\ref{eq:lda2}, \ref{eq:lda3}) vs the dihedral angle number
along the strand, $n$, 
obtained from Monte Carlo simulations. Data are related to systems 
of size $M=15$ within potential energy model $B$.
Different lines correspond (from top to bottom) to tapes with chiral 
equilibrium parameter $\tau_0=0.0,0.1,0.2,0.3$
(Eqs. \ref{eq:tau1} and \ref{eq:tau2}).
\end{quotation}

\noindent{\large\bf Fig. 3. \ }
\begin{quotation}\noindent
Averaged structures obtained from Monte Carlo simulations 
(over the last $10^{6}$ Monte Carlo sweeps) for systems
of size $M=15$ within the potential energy model $A$. 
Here the values of the chirality parameter were $\tau_0=0.25,\
0.5,\ 0.75,\ 1$ respectively.
\end{quotation}

\noindent{\large\bf Fig. 4. \ }
\begin{quotation}\noindent
Averaged structures obtained from Monte Carlo simulations 
(over the last $10^{6}$ Monte Carlo sweeps) for systems
of size $M=45$ within potential energy model $B$. 
$(a)$ Achiral system with $\tau_0 = 0.0$. 
$(b)$ Introduction of chirality in the force field 
($\tau_0 = 0.1$) leads to the stabilization of a 
regularly--twisted supramolecular tape.
$(c)$ A larger twist is obtained for $\tau_0 = 0.3$. 
\end{quotation}

\noindent{\large\bf Fig. 5. \ }
\begin{quotation}\noindent
Histograms of the pitch wave number $k$ 
(expressed in degrees for better clarity)
obtained from Monte Carlo simulations for systems
with size $M=7,\ 15,\ 45$ (from top to bottom).
These data relate to the potential energy model $A$ with $\tau_{0}=1.0$. 
Similar results have been obtained for potential energy model $B$ also.
\end{quotation}


\newpage

\hoffset -15mm
\voffset 10mm

\newpage

\hoffset -15mm
\voffset 40mm

$\left.\right.$\\$\left.\right.$\\$\left.\right.$\\$\left.\right.$\\

\newpage

\hoffset -15mm
\voffset -20mm

\newpage

\hoffset -15mm
\voffset -22mm

\newpage

\hoffset -15mm
\voffset 0mm

\begin{thebibliography}{99}
\bibitem{footnote1}
These positions were taken 
by averaging over the equilibrium Monte Carlo trajectories.
\end{thebibliography}
\end{document}